\documentclass[preprint2]{aastex}

\usepackage{lscape}
\usepackage{apjfonts}

\begin{document}


\title{Lunar Occultations of Eighteen Stellar Sources \\
from the 2.4-m Thai National Telescope
}


\author{A. Richichi\altaffilmark{1}, 
O. Tasuya\altaffilmark{2}, 
P. Irawati\altaffilmark{1}, 
B. Soonthornthum\altaffilmark{1}, 
V.S. Dhillon\altaffilmark{3,4}, 
T.R. Marsh\altaffilmark{5} 
}


\altaffiltext{1}{National Astronomical Research Institute of Thailand, 191 Siriphanich Bldg., 
Huay Kaew Rd., Suthep, Muang, Chiang Mai  50200, Thailand}
\altaffiltext{2}{Department of Physics and Materials Science, Faculty of Science, 
Chiang Mai University, Chiang Mai 50200, Thailand}
\altaffiltext{3}{Department of Physics and Astronomy, University of Sheffield, 
Sheffield S3 7RH, UK}
\altaffiltext{4}{Instituto de Astrofisica de Canarias, 38205 La Laguna, Santa Cruz de Tenerife, Spain}
\altaffiltext{5}{Department of Physics, University of Warwick, 
Gibbet Hill Road, Coventry CV4 7AL, UK}

\email{andrea4work@gmail.com}


\begin{abstract}
We report further results from the program of lunar occultation (LO) 
observations started at the 
2.4-m Thai National Telescope (TNT) in 2014. We have recorded
LO events of 18 stellar sources, leading to the detection of 
four angular diameters and  two binary stars.
With two exceptions, these are first-time
determinations. We could resolve angular diameters
as small as 2 milliarcseconds (mas) and projected separations
as small as 4\,mas.
We discuss the individual results, in the context
of previous observations when available. The first-time
angular diameters for $\omicron$~Psc,
HR~6196 and 75~Leo are in good agreement with expected
values, while that of $\pi$~Leo agrees with the average of previous
determinations but has a higher accuracy.
We find a new secondary in $\omicron$~Psc, as previously
suspected from Hipparcos data.
We also obtain an accurate measurement of the
companion in 31~Ari, revealing inconsistencies in the
currently available orbital parameters.
The TNT, equipped with the fast 
ULTRASPEC  imager, is
the leading facility in Southeast Asia
for high time resolution observations. 
The LO technique at this telescope
achieves a
sensitivity of $i' \approx 10$\,mag, with a potential to detect
several hundreds of LO events per year.
\end{abstract}


\keywords{techniques: high angular resolution -- 
occultations --
binaries: general --
stars: fundamental parameters
}



\section{Introduction}
In a previous paper \citep[][R14 hereafter]{2014AJ....148..100R}
we have described the novel combination of the ULTRASPEC instrument
installed at the 2.4-m Thai National Telescope (TNT), and operated
in drift mode to achieve fast imaging at the level of few milliseconds.
This capability, entirely new for the given longitude range and 
at a telescope which is the largest in the
southeast-asian region, has enabled  a program of
routine lunar occultation (LO) observations. The aim is to
measure angular diameters, to detect and characterize circumstellar
components, and to investigate binary or multiple stars. The achieved
angular resolution is at the milliarcsecond level (mas) and the sensitivity
 $i' \approx 10$\,mag.

This is currently the only routine program of its kind around
the world. Another large program was active at the ESO
Very Large Telescope in the previous decade and produced a 
large number of results
\citep[][and references therein]{2014AJ....147...57R},
but was terminated with the
decommissioning of the ISAAC instrument.
Only few observatories 
around the world remain  suitably
equipped to observe LO, and none of them has a comparable program in place.
Clearly, the high time resolution of this facility can be
employed to observe 
 other classes of objects as well, from cataclysmic variables to
transits, and many kinds of transient phenomena.

We report here on LO results recorded in the
second TNT observing cycle, between December 2014 and May 2015.

\section{Observations and Data Analysis}\label{obs_data}
The observations and the data analysis follow closely what
was already described in R14, and we provide here only
a brief summary. An in-depth general description of ULTRASPEC at TNT
has been presented by \cite{dhillon2014}. For LO, we use the
so-called drift mode of the instrument, in which most of the
detector is masked and light is recorded in two small
square sub-windows. Only one
of the two is effectively used. This allows us to record
uninterrupted continuous sequences, with typical sampling
times around 6~ms and minimal overheads.

For the observations reported here, we have used standard SDSS $r', i', z'$
broad-band filters, as well as two narrow-band
filters, R$_{\rm cont}$ and N86. They have central wavelengths of 6010 and 8611 \AA{}
and full-width half-maxima (FWHM) of 118 and 122 \AA{}, respectively.
Narrow-band filters have the advantage of increasing the contrast
of the diffraction fringes, as well of course of reducing the risk of
saturation in case of strong lunar background.

We typically record about 30\,s data around the predicted time
of the event, to account for possible uncertainties, e.g. due
to proper motions and limb deviations. The sub-windows can be further
rebinned to improve the time resolution -- see the parameters
Sub and Reb
in Table~\ref{table:occlist}.

The result is a sequence of several thousand
frames, which are converted to a FITS cube and analyzed with specifically
developed software \citep[see][and references therein]{2014AJ....147...57R}. 
A crucial step is to adopt an extraction mask
tailored to measure  only the pixels
effectively exposed to the stellar light, thus greatly reducing the
noise contribution from the general background.
Each frame is accurately time-stamped with sub-millisecond precision
thanks to a dedicated GPS signal. 
Time-stamping is also
relevant for other applications such as lunar
limb profiling, which however we do not discuss here.

Note that we concern ourselves only with about 1\,s of data
around the occultation event: the LO diffraction pattern
extends in fact over 0.5\,s or less. This corresponds in practice
to about 2$\arcsec$ on the sky. Thus we are not sensitive to, e.g., 
wide companions.

We use two approaches to analyze the data:
a model-independent maximum-likelihood
method to derive the brightness profile of the
source \citep{1989A&A...226..366R}, and a model-dependent least-squares method
to derive parameters such as angular diameters and
binary separations \citep{richichi96}. The software
includes a number of features, e.g. to derive
upper limits for unresolved sources or to model
low-frequency scintillation. Details are provided
in the references above.

\section{Results}
The  LO events are listed in chronological order
in Table~\ref{table:occlist},
which follows the same format as R14. In summary:
D and R refer to disappearances and reappearances, respectively;
the magnitudes and
spectra are quoted from Simbad; 
the filters were described in Sect.~\ref{obs_data};
Sub and Bin list, respectively,
the size of the detector sub-array
and the on-chip rebinning - i.e., Sub 16x16 and Bin 2x2
lead to frames of size 8x8;
$\tau$  and $\Delta$T 
are  the integration and sampling times, respectively;
S/N is the signal-to-noise ratio, measured as the unocculted stellar
signal divided by the rms of the fit residuals;
and finally UR, Diam, and Bin denote
unresolved, resolved diameter, and binary star, respectively.

\begin{deluxetable}{rrclclcccccrl}
\tabletypesize{\small}
\tablecaption{List of observed events\label{table:occlist}}

\tablehead{
\colhead{Date}&
\colhead{Time}&
\colhead{Type}&
\colhead{Source}&
\colhead{$V$}&\colhead{Sp}&
\colhead{Filter}&\colhead{Sub}&
\colhead{Bin}&\colhead{$\tau$}&\colhead{$\Delta$T}&
\colhead{S/N}&
\colhead{Notes}\\
\multicolumn{2}{c}{(UT)}   & & 
   			 & \multicolumn{1}{c}{(mag)} & 
   			 &  & 
   			\multicolumn{2}{c}{(pixels)} & 
   			\multicolumn{2}{c}{(ms)} & 
   			 & 
}
\startdata
09-Dec-14	&	22:10	&	R	&	SAO 97302	&	8.0	&	G5V	&	N86	&	16x16	&	2x2	&	6.6	&	6.9	&	5.0	&	UR	\\
09-Dec-14	&	22:13	&	R	&	SAO 97303	&	8.7	&	K0V	&	N86	&	16x16	&	2x2	&	6.6	&	6.9	&	2.6	&	UR	\\
02-Jan-15	&	14:53	&	D	&	IRC +20090	&	9.2	&	K0	&	N86	&	16x16	&	2x2	&	6.6	&	6.9	&	11.9	&	UR	\\
26-Jan-15	&	14:14	&	D	&	$\omicron$ Psc	&	4.3	&	G8III	&	N86	&	16x16	&	2x2	&	6.1	&	6.3	&	60.9	&	Bin, Diam	\\
27-Jan-15	&	11:49	&	D	&	31~Ari	&	5.7	&	F7V	&	R$_{\rm cont}$	&	16x16	&	2x2	&	6.6	&	6.9	&	40.1	&	Bin	\\
12-Mar-15	&	20:08	&	R	&	HR 6196	&	4.9	&	G8II/III	&	N86	&	16x16	&	2x2	&	6.6	&	6.9	&	100.8	&	Diam	\\
12-Mar-15	&	20:27	&	R	&	SAO 160044	&	6.7	&	A3III	&	$i^\prime$	&	16x16	&	2x2	&	6.6	&	6.9	&	85.6	&	UR	\\
28-Mar-15	&	14:50	&	D	&	SAO 96977	&	9.3	&	G0V	&	$z^\prime$	&	32x32	&	2x2	&	12.4	&	12.9	&	6.6	&	UR, Wide Bin	\\
28-Mar-15	&	14:50	&	D	&	SAO 96978	&	9.4	&	G0V	&	$z^\prime$	&	32x32	&	2x2	&	12.4	&	12.9	&	6.0	&	UR, Wide Bin \\
30-Mar-15	&	14:56	&	D	&	SAO 98400	&	6.5	&	F2Vp	&	$z^\prime$	&	16x16	&	2x2	&	6.6	&	6.9	&	37.2	&	UR	\\
31-Mar-15	&	16:00	&	D	&	HR 3938	&	6.0	&	K3III	&	N86	&	16x16	&	2x2	&	6.6	&	6.9	&	52.7	&	UR	\\
31-Mar-15	&	17:47	&	D	&	$\pi$ Leo	&	4.7	&	M2IIIab	&	N86	&	8x8	&	no	&	6.1	&	6.3	&	146.6	&	Diam	\\
24-Apr-15	&	14:33	&	D	&	SAO 96634	&	9.0	&	A2V	&	$r^\prime$	&	32x32	&	no	&	25.1	&	25.6	&	21.8	&	UR	\\
29-Apr-15	&	17:32	&	D	&	75 Leo	&	5.2	&	M0III	&	N86	&	16x16	&	no	&	11.9	&	12.2	&	266.9	&	Diam	\\
30-Apr-15	&	17:36	&	D	&	SAO 138554	&	8.5	&	K2	&	$z^\prime$	&	16x16	&	2x2	&	6.6	&	6.9	&	21.3	&	UR	\\
07-May-15	&	19:10	&	R	&	SAO 161004	&	7.5	&	O9.2IV	&	N86	&	16x16	&	no	&	11.9	&	12.2	&	4.3	&	UR	\\
08-May-15	&	17:55	&	R	&	IRC $-$20529	&	9.4	&	M4	&	N86	&	16x16	&	2x2	&	6.6	&	6.9	&	32.1	&	UR	\\
11-May-15	&	20:55	&	R	&	SAO 164693	&	7.5	&	G0	&	$z^\prime$	&	8x8	&	no	&	6.1	&	6.3	&	61.6	&	UR	\\
\enddata
\end{deluxetable}

The stars with a positive determination (i.e., not unresolved)
are listed in Table~\ref{tab:results2}, which also follows a
format used in our previous papers.
Columns 2 and 3 list the measured rate of the event V
and its deviation from the predicted rate V$_{\rm t}$.
This deviation is due mainly to slopes in the local lunar limb $\psi$,
which  can thus be retrieved and are listed in Column 4.
All the deviations and limb slopes are within the norm, based
on the experience of several thousands LO events.
Columns 5 and 6 list the Position Angle and the Contact Angle
of the event, with the limb slope already included.
For the sources found to be resolved, column 7 lists the
best-fitting angular diameter in the uniform disk approximation.
This is strictly valid for the observed wavelength only; conversions to limb-darkened
values can be derived depending on  models.
For the binary stars, columns 8 and 9 list the projected separation
(along the PA direction) and the brightness ratio, respectively.

In the following we discuss in some detail our results, also in the context
of available previous studies. We show figures only for one case
of a resolved angular diameter (Fig.~\ref{fig:hr3950}) and for
one case of a binary star (Fig.~\ref{fig:31ari}).


\begin{deluxetable}{lcrrrrrrr}
\tabletypesize{\small}
\tablecaption{Summary of results: angular sizes (top) and binaries (bottom)\label{tab:results2}}

\tablehead{
\colhead{(1)}&
\colhead{(2)}&
\colhead{(3)}&
\colhead{(4)}&
\colhead{(5)}&
\colhead{(6)}&
\colhead{(7)}&
\colhead{(8)}&
\colhead{(9)}\\
\colhead{Source}&
\colhead{V (m/ms)}&
\colhead{(V/V$_{\rm{t}}$)--1}&
\colhead{$\psi $($\degr$)}&
\colhead{PA($\degr$)}&
\colhead{CA($\degr$)}&
\colhead{$\phi_{\rm UD}$ (mas)} &
\colhead{Proj.Sep.(mas)} & 
\colhead{Br. Ratio} 
}
\startdata
$\omicron$ Psc	&	0.2138	&	$-$4.6\%	&	-0.8	&	144.2	&	73.8	&	2.17$\pm$0.01	&	&	\\
HR 6196	&	0.4345	&	12.3\%	&	4.9	&	345.9	&	231.1	&	1.85$\pm$0.01	&	&	\\
$\pi$ Leo	&	0.6112	&	$-$7.5\%	&	-11.3	&	121.7	&	27.3	&	4.35$\pm$0.01	&	&	\\
75 Leo	&	0.6695	&	$-$0.8\%	&	-1.5	&	132.5	&	18.5	&	2.94$\pm$0.03	&	& 
\\																
\hline
$\omicron$ Psc	&	0.2138	&	$-$4.6\%	&	-0.8	&	144.2	&	73.8	& &	12.4$\pm$0.4	&	41.1$\pm$0.3	\\
31~Ari	&	0.2730	&	$-$4.9\%	&	-1.5	&	307.5	&	60.5	& &	3.76$\pm$0.02	&	1.52$\pm$0.02	\\
\enddata
\end{deluxetable}

\subsection{$\omicron$ Piscium}\label{omipsc}
This star (HR~510, HIP 8198) is a relatively nearby late-type
giant star (G8, 85\,pc) which has been the
subject of several investigations concerning its possible
binary nature. It was included in a catalog 
of stars whose Tycho and Hipparcos data
showed non-linear behavior with time, indicating possible 
astrometric binaries \citep{2005AJ....129.2420M}. 
As such, it was also included as
possible binary in a catalog of bright multiple systems by
\citet{2008MNRAS.389..869E}. 
\citet{2007AA...464..377F} did not find evidence of
changes in the radial velocity, using a statistical analysis that
was however mostly sensitive to stars with parallax $> 20$\,mas.
\citet{1984PASP...96..105H} obtained speckle observations 
at a 4-m telescope, finding it unresolved with an upper
limit of 30\,mas.
Interestingly, no LO event of this star was
ever reported until now.

We obtained a good quality (S/N=61), first ever LO 
light curve of  $\omicron$~Psc.
Thanks to a relatively large contact angle, the efffective limb
velocity was about 3 times slower than normal, leading to an
increased sampling of the fringes which was crucial in assessing
departures from a point source. The data are consistent with
a resolved angular diameter and in addition with a secondary
component 4.0\,mag dimmer than the primary, 
with a separation of 12\,mas projected along PA=144$\degr$.
Our first-time determination of the angular diameter,
when combined with the Hipparcos distance,
results in 19.8\,R$_{\sun}$. 

The companion, given the magnitude difference, is likely to
be a main sequence star, although more details could be ascertained
only with further observations at different wavelengths. Some constraint
on the actual PA can be inferred from the upper limit on the
separation by \citet{1984PASP...96..105H}. Assuming that their
speckle measurements were sensitive to a companion with such a
brightness ratio, the PA would have to be in the --admittedly broad--
range 78$\degr$ to 210$\degr$. Assuming a combined 1\,M$_{\sun}$, a 
real separation between 1 and 2.6\,AU (converted from our measurement 
and from the speckle upper limit), a circular orbit and no inclination effects, the
minimum period would be between 1 and 4 years. Barring significant
inclinations of the orbital plane on the sky, orbital motion might
be detectable within relatively short periods of time, using i.e.
adaptive optics at a very large telescope at visual wavelengths.

\subsection{HR 6196}\label{hr6196}
This star is a late-type giant (G8II/III) and a bright 
near-infrared source (IRC~$-$20325, $K=2.1$\,mag). No angular diameter
measurements are present in the literature.
\citet{1981AJ.....86.1277E} obtained a LO light curve, finding
the source unresolved. Their data were recorded in the blue with
a small telescope and were presumably noisy.
Another previous LO light curve
is listed in the CHARM2 catalog \citep{CHARM2},
however the result was not published. 
\citet{2001A&A...367..521P} estimated a diameter
of 2.0\,mas.

We recorded a high S/N reappearance light curve  of HR~6196.
The data are best fitted by a resolved source model
of 1.85\,mas diameter, which leads to a $\chi^2$ almost 2.5 times lower than
for a point source. We also revisited the old data
set mentioned in CHARM2, which was obtained with an InSb 
photometer in the $K$ band 
at the Calar Alto 1.2-m telescope on July 16, 1997.
Unfortunately, those data are affected by significant 
scintillation and only lead to an upper limit between  2.7 to 3.4\,mas,
consistent with our result but not very constraining.


\subsection{$\pi$ Leonis}\label{pileo}
\begin{figure}
\includegraphics[angle=-90, width=76mm]{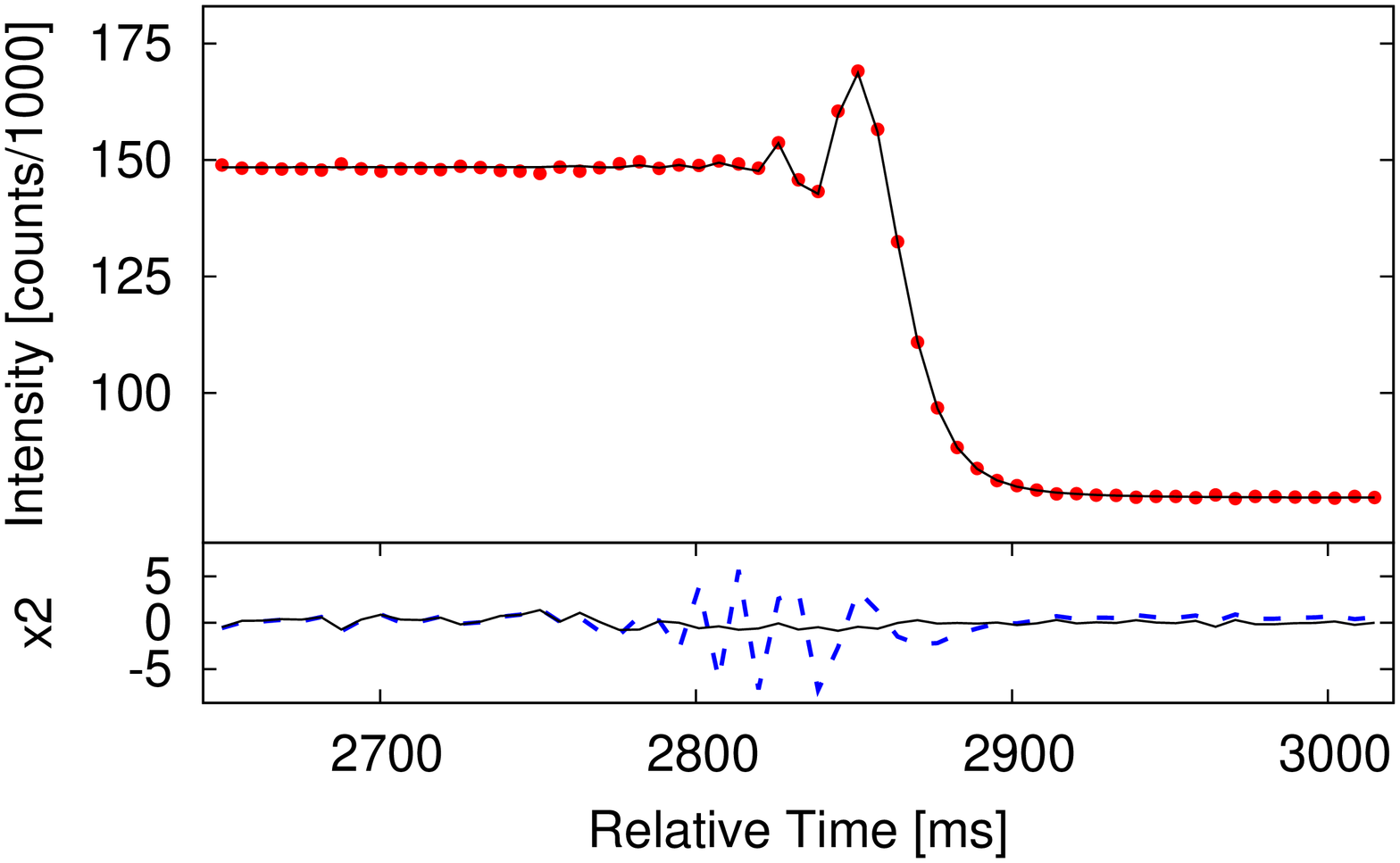}
 \caption{Top panel: occultation data (dots) for 
$\pi$~Leo, and best
 fit by a model with an angular diameter of 4.35\,mas (solid line). 
  The 
lower panel shows the fit residuals, enlarged by a factor of two for clarity, as
a solid line. For comparison, also the residuals of the
best fit by a point-like source are shown (dashed line).
 A color version of this figure is available online.
}\label{fig:hr3950}
\end{figure}
This cool giant (HR~3950, IRC~+10224) has had a number
of spectral classifications, with a general agreement around M2.
It has featured in
many publications especially in the near-IR, where it has
often been used as a calibrator source  due to its
brightness ($K=0.5$\,mag) and minimal variability
\citep[e.g.][]{2004ApJ...605..436M,2008ApJ...673..418W, 2014A&A...572A..17M}.

The angular diameter of $\pi$~Leo has been 
repeatedly determined in the past, with results listed
in the uniform $\phi_{\rm UD}$,
limb-darkened $\phi_{\rm LD}$ or
fully-darkened $\phi_{\rm FD}$ disk hypothesis.
Using the NPOI interferometer
at various wavelengths with an average of 740\,nm,
\citet{1999AJ....118.3032N} reported $\phi_{\rm UD}=4.29\pm0.05$\,mas,
in very good agreement with our own determination.
A large number of LO observations also exist, 
with varying levels of data quality and accuracy.
\citet{1977PASP...89...95V} reported 
$\phi_{\rm LD}=5.9$\,mas in the $R$-band (no error),
while \citet{1978AJ.....83.1100A} reported
$\phi_{\rm FD}=3.9\pm1.0$\,mas in the blue. We note
that in both cases a small sub-meter class telescope
was used, so that scintillation may have played a role.
A "smoothing procedure" was claimed as a possible cause
for the first high value \citep{1978AJ.....83.1100A}.
Later, \citet{1979AJ.....84..247R} reported
$\phi_{\rm UD}=4.58\pm0.33$\,mas, 
while \citet{2013RAA....13.1363B} obtained
$\phi_{\rm UD}=4.9\pm0.5$\,mas, both
in narrow $K$-band filters. 
\citet{2011AJ....141...10S} reported
$\phi_{\rm UD}=3.85\pm0.15$\,mas, using
a 1991 LO light curve obtained in H$_{\alpha}$
with a 1.3-m telescope. The weighted average of
these LO results, neglecting the difference
in wavelengths and limb-darkening, gives
at first approximation $\phi=4.3\pm0.3$\,mas.

We obtained a light curve for $\pi$~Leo
with a quality (S/N=147) better than any previous others.
As shown in Fig.~\ref{fig:hr3950}, the source
is clearly resolved, and our result listed in 
Table~\ref{tab:results2} is consistent with
all good-quality previous determinations, but
with significantly improved accuracy.

\subsection{75~Leonis}\label{75leo}
This is a bright M0 giant (HR~4371), especially luminous
in the near-IR (IRC~+00203, $K=1.4$\,mag) and included
in many publications, some of which covering high-angular
resolution. Despite several attempts, the
star had remained until now essentially unresolved.
These include
speckle observations at  a 4-m telescope with a stated upper limit of 35\,mas
by \citet{1989AJ.....97..510M} and by \citet{1996AJ....112.2260M},
and a three-channel LO light curve from a 1.5-m telescope by 
\citet{1995A&AS..110..107M}.
\citet{1991AN....312..315G} obtained a LO light curve
which was consistent with a companion
at a projected separation of 45\,mas and 0.8\,mag fainter.
However, the data were obtained with a 40-cm telescope in an urban environment
and the authors themselves cautioned about this finding.
Having a stable luminosity, the star has also been included 
in catalogs of spectrophotometric indirect diameter
estimates. For example, \citet{1999AJ....117.1864C} quote
$\phi_{\rm UD}=2.90\pm0.03$\,mas.

Our light curve for 75~Leo has a formally very high
S/N of 266. In fact, the data were taken with a rather
slow sampling (and correspondingly long integration time,
hence the high S/N value), due to
passing clouds that made the star temporarily disappear
during acquisition and forced us to observe in blind mode
with an unusually large sub-window. As a result, the
diffraction fringes are smoothed and poorly sampled.
Nevertheless, these factors can be properly included in the
data analysis and the
$\chi^2$ clearly shows that the star is indeed
resolved. Our resulting angular diameter value is
$\phi_{\rm UD}=2.94\pm0.03$\,mas, in excellent
agreement with the estimate of
\citet{1999AJ....117.1864C}. This represents
the only direct determination
for 75~Leo. 
Barring a very unfavorable combination
of projection directions, 
the bright companion possibly
claimed by \citep[][who did not
mention a PA for their event]{1991AN....312..315G}
should have been seen in our data.

\subsection{31~Ari}\label{sao93022}
\begin{figure}
\includegraphics[angle=-90, width=76mm]{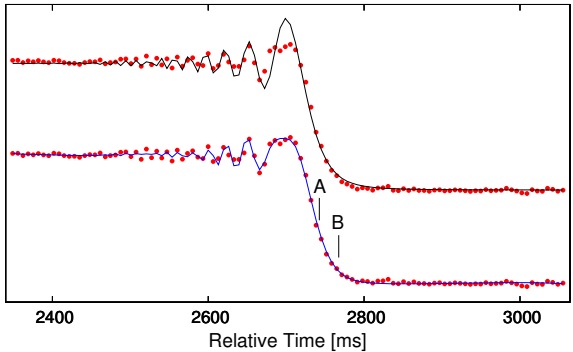}
 \caption{
The occultation data (dots) for 
31~Ari, repeated twice and offset 
for clarity. The top set shows the
best fit by a point-source model, and the bottom
set the best fit by a 
binary model with 3.76\,mas separation. The two segments
mark the times of the geometrical occultation of each component.
The vertical axis is in arbitrary units.
A color version of this figure is available online.
}\label{fig:31ari}
\end{figure}
This bright star (HR~763, SAO 93022) has a spectral type F7V.
Being main sequence, it is as expected relatively nearby:
its Hipparcos parallax places it at 35\,pc. 31~Ari was
first discovered to be binary from a LO by
\citet{1978AJ.....83.1100A}, with projected separation
21\,mas along PA=$265.7$. The magnitude difference
was found to be small both in the blue and in the red,
prompting the authors to speculate similar spectral
types. For comparison with our measurement
(shown in Fig.~\ref{fig:31ari}), they
quoted $\Delta {\rm mag}=0.3\pm0.1$.

Following up on this, 31~Ari was detected as 
binary using speckle by \citet{1982ApJS...49..267M},
and later reported in numerous other publications.
Both \citet{1988PAZh...14..927B} and \citet{1997AJ....114..808M}
computed orbital elements. They are however in considerable
disagreement.
Using the latest, and presumably more complete, of the two we find
that both older 
\citep[e.g.][epochs 1977.75 and 1984.07]{1978AJ.....83.1100A,
2000AJ....119.3084H}
and newer observations
\citep[e.g.][epochs 2010.96 and 2009.75]
{2012AJ....143...42H,  2012AJ....143...10H}
cannot be reproduced. It is thus not surprising that
also our measurement is at odds with the PA and separation predicted
for epoch 2015.07 using the elements by
\citet{1997AJ....114..808M}. Clearly, a reevaluation of all
available measurements for 31~Ari is needed.

\subsection{Other sources}\label{negative}
IRC+20090 was found to have $\phi_{\rm UD}=3.7\pm0.3$\,mas
by \citet{2000MNRAS.317..687T} from a LO event
in the near-IR. The S/N in our light curve
is insufficient to measure the angular diameter.

The two stars SAO~96977 and SAO~96978 are the
components of the wide 
binary system ADS~6146.
We recorded both LO events inside the same sub-window,
about 5\,s apart, and we treated them 
separately. Both were found to be unresolved.

The bright star SAO~98400 (HR~3635) is an intriguing case.
It was claimed as a possible double, seen in two colors
simultaneously, by \citet{1979ApJS...40..475E}. They
mentioned strong noise in their data and did not provide
details about the separation, but they quoted a private
communication from another observer who had
also detected SAO~98400 as binary. Using speckle
at a 4-m telescope,
\citet{1989AJ.....97..510M} detected the companion with
89\,mas separation along PA$=71\degr$. Following that
however, the source was not detected again by speckle in several
attempts including with the russian 6-m telescope
\citep{1993AJ....106.1639M, 1994A&AS..105..503B, 1997AJ....114.1623F}.
The star is at just under 40\,pc distance and one could speculate
a highly eccentric orbit which over the course of a few years
brought the component below the diffraction limits of the
telescopes employed. We note that our LO light curve
had sufficient S/N and sampling rate to detect the
companion if the projected separation had been $\ga 5$\,mas.

SAO~96634 was reported as binary from visual micrometer
observations by \citet{1966JO.....49..341C, 1975A&AS...20..391C},
with separation 71\,mas and PA$=43\fdg8$, unchanged over
nine years. The system was rediscovered by
\citet[apparently unaware of Couteau's work]{1978AJ.....83.1100A}
who recorded two LO events, each in two colors. The projected
separations ($0\farcs3$ and $0\farcs2$) 
however did not agree with Couteau's results, and
also not between  themselves. 
Finally, \citet{1996AJ....112.2260M} could not
resolve the system by speckle observations at a 3.6-m telescope.
Our data are consistent with a companion with 60\,ms projected
separation along PA$=85\degr$, in approximate agreement with
Couteau's values.
However, we do not list this among our binary detections because
there is no significant $\chi^2$ improvement over a
point-source model. The companion should be  detectable by
adaptive optics imaging.

\section{Conclusions}
We reported the latest results from the program
of routine LO observations at the 2.4-m Thai
National Telescope. They include 4 angular diameters,
three of which are first-time determinations, and
two binary stars, of which one was not previously known.

The angular diameters of
$\omicron$~Psc, HR~6196, $\pi$~Leo and 75~leo are found
to be in the range 2 to 4\,mas,
with an average accuracy of 0.5\%. Thus, if combined with
accurate bolometric fluxes, our results have the
potential to constrain the effective temperatures
to the 1\% level, or less than 50\,K.
Concerning the binary stars, we detected companions 
with projected separations of about 4 and 12\,mas
in  31~Ari and $\omicron$~Psc, respectively.
The magnitude  differences (in one case as high as
4\,mag) could be measured to the 1\% level.

We also discussed those stars for which an angular
diameter or a companion had been previously published, but
which we found unresolved. 
The case of SAO~98400 merits
attention, as this binary  was detected repeatedly in the
1980's but not in the following decades. 

As stated in R14, the TNT equipped with the ULTRASPEC
EMCCD fast imager can record LO light curves with few
ms sampling times to about $i' \approx 10$\,mag at S/N=1.
For illustration,
we computed future events taking the SAO catalog as an approximate 
proxy for the sample of stars potentially reachable 
by LO at the TNT.
After filtering for 
good circumstances (Sun and Moon elevations, lunar
phase), we find
that over 500 LO events would be well observable during
each dry season period.

\acknowledgments
VSD and TRM acknowledge the support of the Royal Society and the
Leverhulme Trust for the operation of ULTRASPEC at the TNT.
This research made use of the Simbad database,
operated at the CDS, Strasbourg, France.
The data of 28-31 March 2015 were
collected during the NIATW (NARIT International Astronomical
Training Workshop), with the involvement of the participants.

\clearpage

\end{document}